# Solution for a local straight cosmic string in the braneworld gravity

M. C. B. Abdalla[1,a], P. F. Carlesso[1,b], J. M. Hoff da Silva[2,c]

[1] Instituto de Física Teórica, UNESP, Universidade Estadual Paulista, Rua Dr. Bento Teobaldo Ferraz 271, Bloco II, Barra-Funda, Caixa Postal 70532-2, São Paulo, SP 01156-970, Brazil
[2] Departamento de Física e Química, UNESP, Universidade Estadual Paulista, Av. Dr. Ariberto Pereira da Cunha, 333, Guaratinguetá, SP, Brazil

**Abstract** In this work we deal with the spacetime shaped by a straight cosmic string, emerging from local gauge theories, in the braneworld gravity context. We search for physical consequences of string features due to the modified gravitational scenario encoded in the projected gravitational equations. It is shown that cosmic strings in braneworld gravity may present significant differences when compared to the general relativity predictions, since its linear density is modified and the deficit angle produced by the cosmic string is attenuated. Furthermore, the existence of cosmic strings in that scenario requires a strong restriction to the braneworld tension: $\lambda \geq 3 \times 10^{-17}$, in Planck units.

## 1 Introduction

Cosmic strings were predicted in the 1970s by Kibble [1], as a result of phase transitions in the early universe. The breaking of symmetries in field theories that describe the matter content in the universe can produce such structures. The properties of the defects are related to the epoch in which the symmetry has been broken. For instance, the strings produced for a grand unification theory have a linear density much greater than strings produced later on, for an electroweak theory. The cosmic strings can have cosmological and astrophysical consequences that were exhaustively studied in the literature [2,3].

In spite of the theoretical advances in the cosmic string studies, till now there is no convincing observational evidence as to the existence of such structures. Furthermore, the linear density associated with cosmic strings is strongly bounded by data analysis from the WMAP and most recently from the Planck collaboration [4].

Relatively recent work has shown that cosmic strings can emerge from string theories [5–7]. The production of cosmic fundamental and gauge strings is predicted by braneworld inflation scenarios. These scenarios intend to explain inflation from first principles. The fundamental strings have similar cosmological properties with the gauge strings. For that, a necessary condition is that the fundamental string scale needs to be comparable to the GUT scale of the gauge strings [6].

The close relation between cosmic strings and brane models impels us to study a scenario with cosmic strings within the braneworld gravity context, providing a way to distinguish ordinary cosmic strings predicted from standard cosmology to strings emerging from braneworld scenarios. In fact, in braneworld models the effective gravitational dynamics can suffer modifications when compared with 4-dimensional general relativity (GR) [8–11].

The main aim of this work is to investigate the consequences of braneworld modified gravity equations related to cosmic strings. For this purpose, we analyze the spacetime generated by a cosmic string in an effective braneworld gravitational field. In that scenario, we have a 3-brane (in which the standard model fields are confined or localized) embedded in a five-dimensional bulk, with the extra dimension provided with a $Z_2$ symmetry. The gravitational field equations differ from the usual Einstein's equations due to the presence of two new terms: a term quadratically dependent to the brane stress-energy tensor and a term resulting from the projection of the bulk Weyl curvature tensor on the brane. Therewith, we analyze and discuss the consequences of these two additional terms in a spacetime generated by a straight cosmic string.

The work is organized as follows: in Sect. 2 we introduce the basic gravitational equations of the braneworld model, namely the Randall–Sundrum II model [8]. The gravitational

[a] e-mail: mabdalla@ift.unesp.br
[b] e-mail: pablofisico@ift.unesp.br
[c] e-mail: hoff@feg.unesp.br

Springer

modifications of the brane model with respect to the standard general relativity gravity are then presented. The modifications are exposed via the projective formalism introduced in Refs. [9,10]. In Sect. 3 we introduce the cosmic string model, which is a finite thickness straight cosmic string [12,13]. We obtain an internal solution to the cosmic string in the braneworld model. It allows us to determine the cosmic string effective linear density. Next, we obtain an external solution to the cylindrically symmetric vacuum outside the cosmic string. We link these two solutions by means of junction conditions. The external case presents conical geometry, in which the deficit angle produced by the cosmic strings in the braneworld model is presented. We show that the study of these types of cosmic string allows us to constrain significantly the brane tension. In Sect. 4 we draw our conclusions.

## 2 The braneworld model

The scenario considered here is based on the Randall–Sundrum model [8], which consists of a 3-brane embedded in a five-dimensional bulk. The standard model particles are assumed to live on the brane, whereas the gravity law covers all the five dimensions. The extra spatial dimension is provided with $\mathbb{Z}_2$ symmetry.

The braneworld model implies modifications in the brane gravitational dynamics. Such modifications can be obtained in a perturbative regime, as done in the original paper [8]. Nevertheless it is possible to reach the exact brane effective gravitation through notions of extrinsic geometry and the use of some appropriate junction conditions [9,10], allowing one to fully characterize the modifications coming from the braneworld set-up. The effective gravitational equations are [9,10]

$$G_{\mu\nu} = -\Lambda g_{\mu\nu} + 8\pi G T_{\mu\nu} + k_5^4 \Pi_{\mu\nu} - E_{\mu\nu}, \quad (1)$$

where $G_{\mu\nu}$ is the brane Einstein tensor, $E_{\mu\nu}$ is a term resulting from the projection of the bulk Weyl curvature tensor on the brane, and $\Pi_{\mu\nu}$ is a term related to the brane stress-energy tensor $T_{\mu\nu}$:

$$\Pi_{\mu\nu} = -\frac{1}{4} T_\mu^\sigma T_{\nu\sigma} + \frac{1}{12} T T_{\mu\nu} + \frac{1}{8} g_{\mu\nu} T_{\alpha\beta} T^{\alpha\beta} - \frac{1}{24} g_{\mu\nu} T^2. \quad (2)$$

Equations (1) and (2) are obtained considering the minimal extension of GR to five dimensions, where $k_5^2$ represents the five-dimensional coupling constant. The brane properties are represented by its tension $\lambda$. The Newton gravitational constant is given by those parameters,

$$G = \frac{k_5^4 \lambda}{48\pi}. \quad (3)$$

Furthermore, we have a brane effective cosmological constant given by

$$\Lambda = \frac{1}{2} k_5^2 \left( \Lambda_5 + \frac{1}{6} k_5^2 \lambda^2 \right), \quad (4)$$

where $\Lambda_5$ represents the bulk cosmological constant.

Moreover, there are two equations representing the energy conservation in the bulk–brane system,

$$\nabla^\mu T_{\mu\nu} = 0, \quad (5)$$
$$\nabla^\mu (k_5^4 \Pi_{\mu\nu} - E_{\mu\nu}) = 0. \quad (6)$$

Turning back to the $E_{\mu\nu}$ term, it carries information as regards the bulk curvature, which is unknown. A useful way to express its degrees of freedom is in a cosmological fluid form [11], known as a *Weyl fluid*,

$$E_{\mu\nu} = -k^4 \left[ U \left( u_\mu u_\nu - \frac{1}{3} h_{\mu\nu} \right) + P_{\mu\nu} + Q_{(\mu} u_{\nu)} \right], \quad (7)$$

where we decompose the metric tensor by means of a 4-velocity field ($g_{\mu\nu} = h_{\mu\nu} + u_\mu u_\nu$). $U = -k^{-4} E_{\mu\nu} u^\mu u^\nu$ represents the dark radiation component, $P_{\mu\nu} = -k^{-4} [h^\alpha_{(\mu} h^\beta_{\nu)} - \frac{1}{3} h_{\mu\nu} h^{\alpha\beta}] E_{\alpha\beta}$ is the anisotropic pressure, and $Q_\mu = -k^{-4} h^\alpha_\mu E_{\alpha\beta} u^\beta$ is the energy flux associated to the Weyl (or dark) fluid. In what follows we shall investigate the behavior of a straight cosmic string under the influence of a typical $E_{\mu\nu}$. In Ref. [14] a recent study taking into account the physical effects of dark radiation in the bulk as well as on the brane is presented.

## 3 The solution for a straight cosmic string

In order to obtain a gravitational solution for spacetime shaped by the presence of the cosmic string we need to define the stress-energy tensor of the source. Local gauge strings have the energy distribution strongly localized over the symmetry axis [15]; this is true for cosmic string solutions both in Minkowski and curved spacetimes [16]. In addition, if we consider a straight cosmic string, a suitable representation for its stress-energy tensor is given by [12,13]

$$T_\mu^\nu = -\epsilon \, \text{diag}(1, 0, 0, 1), \quad (8)$$

in cylindrical coordinates $\{t, \rho, \phi, z\}$. In this representation, the cosmic string is characterized by two constants: $\rho_0$ representing the string radius and $\epsilon$ representing the energy density. For $\rho < \rho_0$ the energy density assumes a constant value $\epsilon$, whereas for $\rho > \rho_0$ it vanishes. Thus we can compute two solutions for the string spacetime: an internal constant energy solution and an external vacuum solution. These solutions will be related by means of some junction conditions on the border of the string. In the usual general relativity, it implies a conical spacetime surrounding the string in which the deficit angle is related to the string linear density [12,13].

GR solutions for straight cosmic strings lead to a conical spacetime [2], with an exterior line element given by

$$ds^2 = -dt^2 + dz^2 + dr^2 + (1 - 8G\mu)r^2 d\phi^2.$$

The constant energy approximation (8) leads to the same GR solution, but it does not generate a singularity at the origin, so it is a good approximation for straight cosmic strings. Although the solution (9) can generate gravitational lensing and contribute to overdensities in the matter distribution due to the conical geometry, it is a flat solution which implies no gravitational force due to the cosmic string. A more realistic model needs to take into account a wiggly structure for the cosmic string.

### 3.1 The internal solution

We proceed to obtain a solution for the above cosmic string in braneworld effective gravity. The extra terms in effective equations can break the cylindrical symmetry and spacetime staticity. In order, however, to compare with the GR solution for the same system, we will seek a static and cylindrically symmetric solution. These assumptions imply the following form for the spacetime line element:

$$ds^2 = -e^{2\alpha(\rho)}dt^2 + e^{2\beta(\rho)}d\rho^2 + e^{2\gamma(\rho)}d\phi^2 + e^{2\delta(\rho)}dz^2, \quad (9)$$

which allows us to compute the left-hand side of Eq. (1). Furthermore, considering the stress-energy tensor (8) the expression (2) and $\Pi_{\mu\nu}$ lead to only two non-vanishing components:

$$\Pi^1_1 = \Pi^2_2 = \frac{1}{12}\epsilon^2. \quad (10)$$

Since the influence of the cosmological constant is not relevant for our analysis, we discard it in the effective equations (1). The remaining term is the Weyl fluid. In order to accomplish a solution for the static and cylindrically symmetric metric, we discard the energy flux term in (7). It allows us to write the Weyl fluid in the following form:

$$E^\nu_\mu = -k^4 \text{diag}\left[-U, P_1 + \frac{1}{3}U, P_2 + \frac{1}{3}U, P_3 + \frac{1}{3}U\right], \quad (11)$$

where $U$ and $P_i$ are functions of the radial coordinate only. Furthermore, the $P_i$ functions satisfy a null-trace condition

$$P_1 + P_2 + P_3 = 0. \quad (12)$$

We stress that Eq. (11) cannot be applied to every braneworld context. Indeed, depending on the spacetime symmetries the Weyl tensor vanishes identically. However, for braneworld models with some extra gravitational influence in the bulk, as a bulk black hole for instance [17], or the existence of discrete gravitational modes endowed by the presence of a second brane in the bulk, the Weyl tensor projection can be explored in a broader sense [11]. Equations (8), (10), and (11) allow us to compute the conservation equations. From Eq. (5) we have

$$\alpha' + \delta' = 0, \quad (13)$$

which implies $\alpha + \delta = $ constant. The constant can be absorbed by a simple coordinate change, leading to

$$\delta = -\alpha. \quad (14)$$

Equation (6) leads to

$$\frac{1}{3}U' + P'_1 + \frac{4}{3}\alpha'U + (\alpha' + \gamma' + \delta')P - \gamma'P_2 - \delta'P_3 = 0. \quad (15)$$

Furthermore, we can compute the gravitational equations from (1),

$$e^{-2\beta}(\gamma'^2 + \alpha'^2 - \beta'\gamma' + \beta'\alpha' - \gamma'\alpha' + \gamma'' - \alpha'') = -8\pi G\epsilon - k^4 U, \quad (16)$$

$$-e^{-2\beta}\alpha'^2 = \frac{k_5^4 \epsilon^2}{12} + \frac{k^4}{3}U + k^4 P_1, \quad (17)$$

$$e^{-2\beta}\alpha'^2 = \frac{k_5^4 \epsilon^2}{12} + \frac{k^4}{3}U + k^4 P_2, \quad (18)$$

$$e^{-2\beta}(\alpha'^2 + \gamma'^2 + \alpha'\gamma' - \alpha'\beta' - \beta'\gamma' + \alpha'' + \gamma'') = -8\pi G\epsilon + \frac{k^4}{3}U + k^4 P_3, \quad (19)$$

in which we already imposed the constraint (14).

Subtracting Eq. (17) from (18) we have

$$2e^{-2\beta}\alpha'^2 = k^4(P_2 - P_1). \quad (20)$$

Lorentz invariance in the $z$ direction requires $\alpha = \delta$, but it implies, from (14), $\alpha = 0$. So, the above equation tells us that Lorentz invariance requires $P_1 = P_2$. In order to compare the case at hand with the usual GR solutions, we assume that this is indeed the case. Therefore, we obtain relations for $U$ and $P_3$,

$$U = -\frac{k_5^4 \epsilon^2}{12k^4} \quad \text{and} \quad P_3 = \frac{k_5^4 \epsilon^2}{9k^4}. \quad (21)$$

Moreover, $P_1$ and $P_2$ are determined by $P_3$, through the traceless condition (12). These constraints turn Eqs. (16) and (19)

to the same expression:

$$e^{-2\beta}(\gamma'^2 - \beta'\gamma' + \gamma'') = -8\pi G\epsilon + \frac{k_5^4 \epsilon^2}{12}. \tag{22}$$

Now, we have two variables and one equation to solve, therefore it is necessary to impose an additional constraint. Fortunately, it is possible to link two of these variables by choosing a specific gauge [18], allowing for the imposition $\beta = \delta$. It implies

$$\gamma'^2 + \gamma'' = -8\pi G\epsilon + \frac{k_5^4 \epsilon^2}{12}. \tag{23}$$

Furthermore, the above assumption satisfies automatically the conservation equation (15). The general solution to the above equation is easy to obtain:

$$\gamma(\rho) = \ln\left[C_1 \sin\left(\frac{\rho - C_2}{\rho_*}\right)\right], \tag{24}$$

where $C_1$ and $C_2$ are constants and

$$\rho_* = \frac{1}{\sqrt{8\pi G\epsilon - \frac{k_5^4 \epsilon^2}{12}}}. \tag{25}$$

If we assume a Minkowski spacetime in the limit $\rho \to 0$, it implies $C_1 = \rho_*$ and $C_2 = 0$. In this case, the internal solution is

$$\gamma(\rho) = \ln\left[\rho_* \sin\left(\frac{\rho}{\rho_*}\right)\right], \tag{26}$$

which is very similar to the GR solution for the same kind of string. The difference is due to the presence of a quadratic $\epsilon$ term in Eq. (25).

Once we have determined the internal spacetime of the string, we can compute the effective string linear density, obtained by integration of the string effective energy density over the transverse dimensions $\phi$ and $\rho$:

$$\mu_{\text{eff}} = \int_0^{2\pi} \int_0^{\rho_0} \left[\epsilon + \frac{1}{8\pi G}\left(\frac{1}{12}k_5^4 \epsilon^2\right)\right] \rho_* \sin\left(\frac{\rho}{\rho_*}\right) d\phi d\rho, \tag{27}$$

whose integration leads to

$$\mu_{\text{eff}} = \left(2\pi\epsilon\rho_*^2 + \frac{1}{48G}k_5^4 \epsilon^2 \rho_*^2\right)\left[1 - \cos\left(\frac{\rho_0}{\rho_*}\right)\right]. \tag{28}$$

Gauge strings have many of their properties determined by the scale parameter $\eta$ [19] of the broken gauge theory. The string radius is related to such a scale by

$$\rho_0 \simeq \frac{1}{\eta}, \tag{29}$$

while the string volumetric density is

$$\epsilon \simeq \eta^4, \tag{30}$$

both in Planck units. The $\epsilon$ parameter is estimated taking into account the string radius and considering that the string linear density is nearly the same in both curved and flat space [16]. With these assumptions, we can simplify Eq. (28), while using Eq. (3) we get a relation for the effective linear density which depends on two parameters,

$$\mu_{\text{eff}} = \frac{1}{4G}\left(\frac{1 + \eta^4/2\lambda}{1 - \eta^4/2\lambda}\right)\left[1 - \cos\left(\eta\sqrt{8\pi G\left(1 - \frac{\eta^4}{2\lambda}\right)}\right)\right]. \tag{31}$$

The gravitational and cosmological implications of cosmic strings are intimately related to its linear density. The above relation allows us to implement an observational restriction to the model. Data from the Planck collaboration have restricted the linear density of cosmic strings [4] as

$$\mu_{\text{eff}} \leq 10^{-7}, \tag{32}$$

where we use $G = c = 1$ (Planck units). It constrains the possible $\eta$ and $\lambda$ values. As we can see in Fig. 1, for large values of $\lambda$ we have the restriction $\eta \leq 1.8 \times 10^{-4}$. For lower values of $\lambda$ (which implies the increasing of the braneworld effects) the restriction over $\eta$ becomes more accentuated.

We can specify the previous results for different types of strings. A string generated at grand unification theories (GUT) energy scales has $\eta_{\text{GUT}} \sim 10^{-4}$, while a string generated during an electroweak phase transition has $\eta_{EW} \sim 10^{-17}$ [3]. By fixing these values, it is possible to obtain bounds over the brane tension for each type of string.

As we can see from Fig. 2, for lower values of $\lambda$ the observational constraint is extrapolated. It implies a minimum value to the brane tension: if we assume the existence of GUT-strings in the brane scenario considered here, we need $\lambda \geq 3 \times 10^{-17}$. The same calculation for electroweak strings leads to the constraint $\lambda \geq 2 \times 10^{-95}$. The restriction due to GUT-strings is very strong. In fact, previous attempts to constrain the brane tension by astrophysical measurements have given less restrictive bounds [20,21].

3.2 The external solution

The external solution will determine the gravitational influence of the strings in the spacetime, as lensing effects and structure formation seeds. To obtain an external solution we start with the same assumptions of the internal solution ($\alpha = \delta = \beta$). It will generate a static, cylindrically symmetric (endowed with Lorenz invariance in the $z$ direction for the solution). The external vacuum equations obtained from (1), where $T_{\mu\nu}$ and $\Pi_{\mu\nu}$ vanish and $E_{\mu\nu}$ is given by (11), read

$$e^{-2\alpha}(\gamma'^2 + \gamma'' + \alpha'') = -k^4 U, \tag{33}$$

$$e^{-2\alpha}(2\alpha'\gamma' + \alpha'^2) = \frac{k^4}{3}U + k^4 P_1, \tag{34}$$

$$e^{-2\alpha}(\alpha'^2 + 2\alpha'') = \frac{k^4}{3}U + k^4 P_2, \tag{35}$$

$$e^{-2\alpha}(\gamma'^2 + \gamma'' + \alpha'') = \frac{k^4}{3}U + k^4 P_3. \tag{36}$$

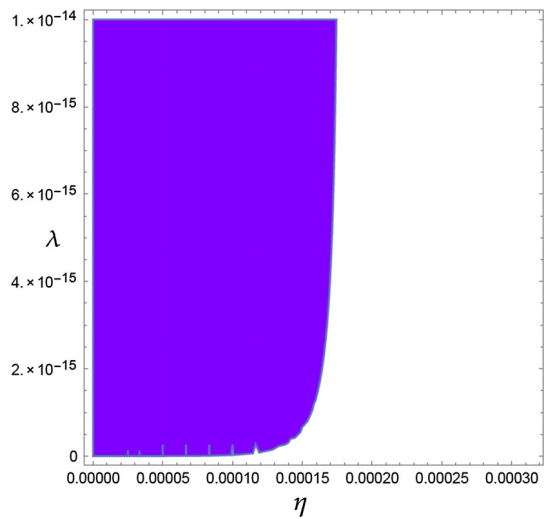

**Fig. 1** Allowed values (*shaded region*) for $\eta$ and $\lambda$ to a cosmic string in braneworld gravity, in Planck units

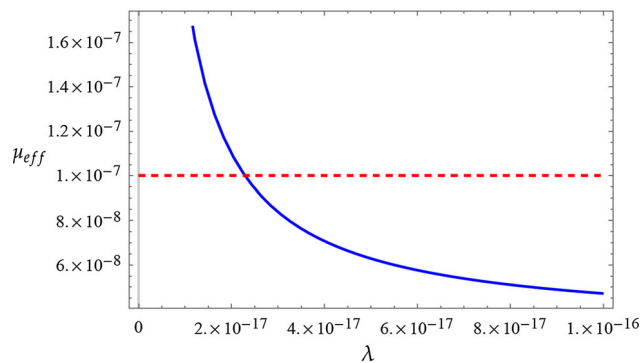

**Fig. 2** String linear density to a GUT-string as a function of $\lambda$, in Planck units. The threshold *dashed line* denotes the observational constraint

Furthermore, the conservation equation (15) becomes

$$\frac{1}{3}U' + P_1' + \frac{4}{3}U\alpha' + P_1(2\alpha' + \gamma') - P_2\gamma' - P_3\alpha' = 0. \tag{37}$$

From Eqs. (33) and (36) we have $P_3 = -\frac{4}{3}U$ and these two equations become identical. Deriving Eq. (34) we obtain an expression for $\frac{1}{3}U' + P_1'$, and by using Eqs. (34) and (35) it is possible to find the expressions for $P_1$ and $P_2$. Replacing these results in (37) we recover (33). Hence the conservation equation does not contribute with additional information over the system of equations and we have three equations to determine four variables. It renders the system undetermined.

We can rewrite Eq. (35) as a function of $P_1$ and $U$:

$$e^{-2\alpha}(\alpha'^2 + 2\alpha'') = \frac{5k^4}{3}U - k^4 P_1. \tag{38}$$

Equation (38), along with (33) and (34), sets up the system of equations to be solved. Unlike the internal solution, it is not easy to find a general solution for the above system of equations. However, we need to impose another restriction over the variables in order to close the system of equations.

In the external case, the vacuum equations differ from GR only due to the $E_{\mu\nu}$ term. It carries information as regards the bulk curvature. We assume here that its influence is very weak, or even negligible when compared with the energy scales of the internal solution, that is, $E_{\mu\nu} \simeq 0$. It may sound as an oversimplification, but we are just disregarding the nonlocal five-dimensional Kaluza–Klein wave effects in the exterior solution when contrasted to the short distance case performed by the interior solution. To the above system of equations, this is equivalent to impose a constant value to $\alpha$. It supplies us with the necessary constraint to solve the system, which leads to only one equation to solve:

$$\gamma'^2 + \gamma'' = 0, \tag{39}$$

whose general solution is

$$\gamma(r) = \ln(ar + b). \tag{40}$$

In the current literature, solutions to the GR case where obtained by assuming $b = 0$. In this case we have the so-called conical spacetime solution [12,13]. The constant $a$ is related to the deficit angle of the spacetime.

### 3.3 The junction conditions

It is necessary to match the internal and external solutions previously obtained. This can be done by imposing the well-known junction conditions [22]. The first one is to require the same values for the metric components in both solutions, on the border ($\rho_0 = r_0$) of the string:

$$g_{\mu\nu}^{(i)}(\rho_0) = g_{\mu\nu}^{(e)}(r_0), \tag{41}$$

Furthermore, the second matching condition says that the extrinsic curvature associated to both solutions needs to be the same on the border of the string,

$$K^{(i)}_{\mu\nu}(\rho_0) = K^{(e)}_{\mu\nu}(r_0), \qquad (42)$$

where $K_{\mu\nu}$ is given by

$$K_{\mu\nu} = h^\alpha_\mu \nabla_\alpha n_\nu. \qquad (43)$$

The above conditions imply the following constraints on the metric variables:

$$\gamma(\rho_0) = \gamma(r_0), \qquad (44)$$
$$\gamma'(\rho_0) = \gamma'(r_0). \qquad (45)$$

For the solutions obtained in the previous sections, it implies a constraint over the deficit angle of the external solution,

$$a = \cos\left(\frac{\rho_0}{\rho_*}\right). \qquad (46)$$

Equation (46) can be written in terms of the mass scale of the string and the brane tension,

$$a = \cos\left(\eta\sqrt{8\pi G(1 - \eta^4/2\lambda)}\right). \qquad (47)$$

Therefore, the brane model predicts a deficit angle ($\Delta = 2\pi(1-a)$) that differs from the usual GR case and the difference depends on the relation between the cosmic string energy density, $\epsilon$, and the $\lambda$ parameter, the brane tension. The GR result is recovered in the $\lambda \to \infty$ limit, as expected. Making use of Eq. (31), the constant $a$ can also be expressed in terms of the effective energy density,

$$a = 1 - 4\mu_{\text{eff}}\left(\frac{1 - \eta^4/2\lambda}{1 + \eta^4/2\lambda}\right), \qquad (48)$$

which is analogous to the result obtained in previous work for the GR case [12,13,23], but here there is a dependence on the brane effective gravity contributions represented by the brane tension $\lambda$.

The modifications with respect to GR become relevant at low values of $\lambda$. In this case, the deficit angle is attenuated, as can be read from Fig. 3. Particularly, for $\lambda = \eta^4/2$, the deficit angle vanishes and it becomes impossible to observe any astrophysical and/or cosmological consequence due to the presence of local straight cosmic strings in the universe.

A particular effect occurs if $\lambda < \eta^4/2$: the expression under the square root of (47) becomes negative and we have an imaginary number; in this case the trigonometric function becomes hyperbolic. Therefore, we have a conical spacetime with negative deficit angle. It means that cosmic strings can produce standard effects as light deflections and density perturbations in the universe, but they do not produce duplicated

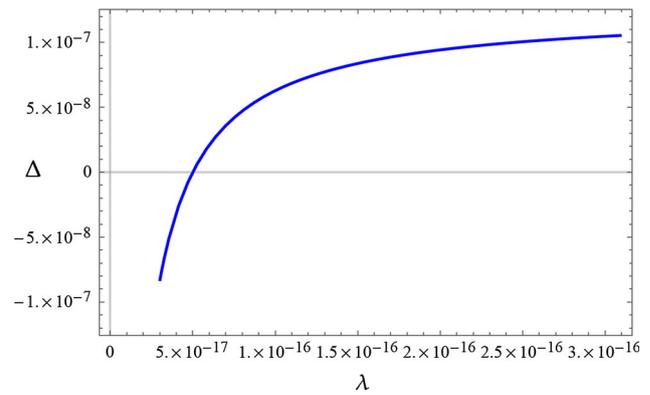

Fig. 3 Graphic of the deficit angle $\Delta$ in units of $2\pi$ as a function of $\lambda$. Notice that for high $\lambda$ values there is no departure from the standard GR deficit angle prediction

images by gravitational lensing, since the spacetime around the cosmic string produces divergent trajectories instead of convergent ones, as in conical spacetimes with a positive deficit angle. Finally, as $\lambda \to 0$ the deficit angle goes to minus infinity, but in view of the $\lambda$ minimum, which is to be proportional to $\eta^4$, arbitrary negative values to the deficit angle are not expected rendering a well-defined $\Delta$ anyway.

## 4 Conclusions

We have seen that the gravitational equations in a braneworld scenario differ from GR by the presence of two additional terms. It produces changes in the solution for the spacetime generated by the cosmic string. We have assumed a cylindrically symmetric and static Weyl fluid. It has allowed us to obtain a solution with such characteristics and then compare it with the GR solution. Yet the solution does not present Lorentz invariance in the $z$ direction. Such a behavior emerges naturally in the GR solution [13]. In order to compare the solutions, we impose by hand the Lorentz symmetry. The result is a spacetime with conical geometry for the external solution and an effective string linear density differing from the Minkowski and GR predictions [2,12].

From the internal solution, it was possible to constrain the braneworld tension in order to satisfy the observational measurements from the Planck collaboration. By assuming the existence of cosmic strings generated by spontaneous symmetry breaking at the early universe, we were able to impose a physical constraint over $\lambda$. The value obtained is much stronger than the previous restrictions obtained in the literature.

Furthermore, we see that the deficit angle of a cosmic string can be significantly attenuated due to the braneworld gravitational corrections. It could explain the negative experimental results in the data of cosmic strings imprints based upon $\Delta$, like lensing effects caused by the conical spacetime generated by the cosmic string. We have shown that a particular value of the brane tension, $\lambda = \eta^4/2$, makes the deficit angle vanishing and eliminates the possibility of lensing effects due to cosmic strings. Such a behavior has also consequences in other cosmological effects of cosmic strings, such as wake formation (overdensity in the matter distribution) by cosmic strings traveling in the universe [24,25], which strongly depends on the deficit angle. These results provide us with a manner to distinguish the effects of cosmic strings in standard cosmology from the braneworld alternative scenarios. Moreover, an eventually negative deficit angle would lead to a different density of created particles per mode, for instance, in a scenario as the one studied in Ref. [26]. The reason is that a negative deficit angle can lead to a slightly different, but relevant, separation constant of the Klein–Gordon equation in the background generated by the cosmic string.

In order to finalize these remarks, we shall mention that the observational restriction on the string linear density is mostly obtained from calculations of cosmic string effects based on GR [4] (model dependent measurements). In view of the results obtained in this work, the brane tension can indeed play a relevant role in cosmic string gravitational effects. Hence we are compelled to explore genuine braneworld effects expecting relevant departures from GR studies.

**Acknowledgments** M. C. B. Abdalla and J. M. Hoff da Silva acknowledge Conselho Nacional de Desenvolvimento Científico e Tecnológico (CNPq) for partial financial support (482043/2011-3; 308623/2012-6). P. F. Carlesso thanks the CAPES-Brazil for supporting his doctorate studies.